\documentstyle[12pt,epsfig]{article}

\def\fp#1#2#3{{Fortschr. Phys. }{\bf #1~}(19#2)~#3}
\def\plb#1#2#3{{Phys. Lett. }{\bf B#1~}(19#2)~#3}
\def\zpc#1#2#3{{Z. Phys. }{\bf C#1~}(19#2)~#3}

\def\prep#1#2#3{{Phys. Rep. }{\bf #1~}(19#2)~#3}
\def\prd#1#2#3{{Phys. Rev. }{\bf D#1~}(19#2)~#3}

\def\cpc#1#2#3{{Comput. Phys. Commun. }{\bf #1~}(19#2)~#3}

\def\epjc#1#2#3{{Eur. Phys. J. }{\bf C#1~}(19#2)~#3}
\def\nima#1#2#3{{Nucl. Instrum. Meth. }{\bf A#1~}(19#2)~#3}

\newcommand{\lsim}{\raisebox{-0.13cm}{~\shortstack{$<$ \\[-0.07cm] $\sim$}}~}
\newcommand{\gsim}{\raisebox{-0.13cm}{~\shortstack{$>$ \\[-0.07cm] $\sim$}}~}

\newcommand{\non}{\nonumber}

\newcommand{\cir}{\mbox{\small${\cal C}$}}
\newcommand{\pta}{\boldmath{\mbox{${\rm P}_t$}}}
\newcommand{\pca}{\boldmath{\mbox{${\rm P}_c$}}}
\newcommand{\ple}{\boldmath{\mbox{$\lambda_e$}}}
\newcommand{\pte}{\boldmath{\mbox{$\zeta_\bot$}}}
\newcommand{\tpta}{\boldmath{\mbox{$\tilde{\rm P}_t$}}}
\newcommand{\tpca}{\boldmath{\mbox{$\tilde{\rm P}_c$}}}
\newcommand{\tple}{\boldmath{\mbox{$\tilde\lambda_e$}}}

\newcommand{\dis}{\displaystyle}
\newcommand{\dfrac}{\dis\frac}
\newcommand{\gaga}{\gamma\gamma}
\newcommand{\mpp}{M_{++}}
\newcommand{\mmm}{M_{--}}
\newcommand{\mpm}{M_{+-}}
\newcommand{\mmp}{M_{-+}}
\newcommand{\re}{{\sf Re}}
\newcommand{\im}{{\sf Im}}
\newcommand{\dd}{{\rm d}}
\newcommand{\dg}{{\rm d}{\Gamma}}

\newcommand{\dsh}{{\rm d}{\hat\sigma}}
\renewcommand{\r}{\rho}
\newcommand{\x}{\xi}
\newcommand{\tx}{\tilde{\xi}}
\newcommand{\beq}{\begin{eqnarray}}
\newcommand{\eeq}{\end{eqnarray}}
\newcommand{\bea}{\begin{array}}
\newcommand{\eea}{\end{array}}
\begin{document}

\renewcommand{\thefootnote}{\fnsymbol{footnote}}

\thispagestyle{empty}
\setcounter{page}{0}

\begin{flushleft}
DESY 99-167
\\
LC-TH-2000-002
\\
{\tt hep-ph/9912467}
\\
December 1999 
\end{flushleft}
\vspace{3cm}

\begin{center}

{\Large \bf
Resonant production of heavy MSSM Higgs bosons\\
\vspace{3mm}
at the Photon Collider
\footnote{
{\it Contribution to the Proceedings of the 2nd Joint ECFA/DESY Workshop, 
``Physics Studies for a Future Linear Collider"
(R. Heuer, F. Richard, P. Zerwas, eds.),
to appear as DESY report 123F.
}
}
}

\vspace{5mm}

{\large Jos\'e I. Illana}\footnote{
On leave from Departamento de F{\'\i}sica Te\'orica y del Cosmos,
Universidad de Granada, Fuentenueva s/n, E-18071 Granada, Spain.\\
\indent E-mail: {\tt jillana@ifh.de}}

\vspace{5mm}

{\sl Deutsches Elektronen-Synchrotron DESY,\\
     Platanenallee 6, D-15738 Zeuthen, Germany}

\end{center}


\begin{abstract}
Assuming a light Higgs boson is discovered, its nature may remain unknown
in case that no supersymmetric particles are found.
The detection and study of heavier Higgs particles is then of great
interest. For this purpose the Compton-collider option of a high energy 
$e^+e^-$ linear collider is optimal both to produce Higgs 
bosons and to reveal their CP-parity. Assuming realistic photon 
luminosities for various configurations of laser and linac polarizations, 
we study the heavy, neutral MSSM Higgs signals for the most relevant 
decay modes as well as their corresponding backgrounds. The MSSM $H$ and $A$ 
Higgs bosons with masses up to a $\sim$80\% of the linac c.m.s. energy may be 
observed and their CP-parity tested for Higgs masses $\lsim$450 GeV, 
at the $\gaga$ mode of a $\lsim$1 TeV linear collider. 
\end{abstract}

\newpage
\section{Introduction}

The global fits to electroweak precision observables provide indirect evidence
of a relatively light Higgs boson, $M_{H^0}=76^{+85}_{-47}$ GeV \cite{LEPEWWG}.
Its discovery is one of the main goals of present and planned experiments.
When such a particle is found it will be of utmost importance to determine
its properties, in particular to know
whether it corresponds or not to the only Higgs boson of the Standard Model 
(SM) $H^0$. Most of the models beyond the SM predict an extended Higgs sector
\cite{higgsreview}. 
In the minimal supersymmetric SM (MSSM) \cite{MSSM} one has
two Higgs doublets yielding five physical Higgs particles after the spontaneous 
breaking of the gauge symmetry: 
two CP-even, neutral Higgs bosons, $h$, which must be light ($M_h\lsim 130$ GeV 
\cite{HWG}) and $H$, which can be heavy; one CP-odd, neutral $A$ and two 
charged Higgses $H^\pm$. The current direct searches provide the 
limits $M_{H^0}>89.7$ GeV, $M_h>79.6$ GeV, $M_A>80.2$ GeV \cite{LEPHWG}.

In case that a light Higgs particle, with mass below $\sim130$ GeV, is 
discovered at LEP 2, Tevatron or at the future LHC or a high energy lepton 
collider, and no supersymmetric particles are observed, it will have very 
probably SM-like properties: the same quantum numbers and almost identical 
decay rates \cite{decoupling}. 
It is then clear that discriminating SM and MSSM will require either 
{\em (i)}~the detection of the slight differences between $H^0$ and $h$, or 
{\em (ii)}~the direct observation of heavier Higgs bosons. 

For both purposes the photon-collider option of a linear $e^+e^-$ collider
\cite{idea} has been proven to be optimal \cite{borden}.
In first place, the best way to discriminate between the SM and the lightest
MSSM Higgs particles is to look at the {\em loop-induced} Higgs boson couplings
such as $\Phi gg$, $\Phi Z\gamma$ and $\Phi\gaga$, with $\Phi=H^0$ or $h$.
They are very sensitive to effects from {\em new physics} and, among them, the 
last one is the most promising. The measurement of the two-photon width of an 
intermediate-mass Higgs boson with an accuracy of $\sim2\%$ is possible 
\cite{soldner}.
Secondly, the Higgs bosons are produced {\em resonantly} by the fusion
of two photons with invariant masses that can be as large as $\sim 80\%$
of the energy of the $e^+e^-$ collider energy. This means, in general, 
a higher discovery reach than using $e^+e^-$ or hadron processes 
at the same energy.

The determination of the CP nature of the Higgs bosons is also an important 
issue. The photon collider provides the ``most elegant way" 
\cite{higgsreview} to determine whether the Higgs field $\varphi$ is a pure or 
a CP-mixed state. Since CP-even and CP-odd components couple with similar 
strength to $\gaga$ (via one-loop graphs), the CP-odd components are not 
masked such as occurs in the processes where the $ZZ\varphi$ or $WW\varphi$ 
couplings are involved.
 
In this work we pursue possibility {\em (ii)}. In the context of a photon 
collider we study the discovery reach and the discrimination power
of the CP-parity of the heavy neutral Higgs bosons of the MSSM.  
The work is organized as follows. The main features of the
photon collider are outlined in Section 2. The $\gaga$ luminosity and 
the polarization effects are introduced in Section 3. The
Higgs-boson phenomenology at the photon collider is presented in Section 4.
The event rates for decaying heavy MSSM Higgs bosons and their backgrounds 
as well as the statistical significances of the signals for the
case of circularly polarized lasers are given in Section 5. In Section 6,
linearly polarized lasers are used to distinguish between $H$ and $A$. 
Finally the main conclusions are summarized in Section 7.

\section{The photon collider}

Due to severe synchrotron radiation in storage rings, future $e^+e^-$
colliders in the TeV region will be linear. Unlike the situation in
storage rings, in linear colliders each bunch is used only once.
This enables to use of electrons or positrons for the production
of high energy photons to obtain colliding $\gaga$ and $\gamma e$ beams
\cite{idea}. The $e\to\gamma$ conversion mechanism in the photon linear
collider (PLC) is the Compton backscattering of electron and laser photos. 
This mechanism presents advantages over others (bremsstrahlung, beamstrahlung) 
because of the possibility to obtain high energy real photons, high 
monochromaticity and a high degree of polarization of backscattered photons 
(using polarized lasers and electron beams) and good background conditions.

\subsection{Spectrum of backscattered photons}

The Compton scattering of low energy photons ($\omega_0\sim$ 1 eV) by high 
energy electrons ($E_b\sim$ 100 GeV) results in a tight bunch of backscattered 
photons.
The kinematics of the process is governed by the dimensionless parameter $x_0$,
defined from $W^2_{e\gamma}\simeq m^2_e(x_0+1)$, $x_0\equiv4\omega_0E_b/m^2_e$. 
The maximal energy fraction $y=\omega/E_b$ carried by the backscattered photons 
is $y_{\rm max}=x_0/(x_0+1)$ and their scattering angle is $\theta(y)\approx
\theta_0\sqrt{y_{\rm max}/y-1}$, with $\theta_0\equiv m_e\sqrt{x_0+1}/E_b$, 
of the order of a few microradians.

The spectrum of the scattered photons $f_C(y)$ depends on the 
product of the mean helicity of the initial electrons and the degree of 
circular polarization of the laser photons, $\ple\pca$, \cite{NIM219}
\beq
\dfrac{\dd f_C}{\dd y}=\dfrac{1}{\sigma_C}\dfrac{\dd\sigma_C}{\dd y}
	=\dfrac{1}{\sigma_C}\dfrac{2\sigma_0}{x_0}
        \left[\dfrac{1}{1-y}+1-y-4r(1-r)-2\ \ple\pca\ x_0 r
                (2r-1)(2-y)\right] ,
\label{eq2}
\eeq
where $\sigma_0\equiv\pi\alpha^2/m^2_e$, $r\equiv y/x_0(1-y)$ and the total 
Compton cross-section is
\beq
\sigma_C&=&\sigma_C^{\rm np}+2{\ple\pca}\ \sigma_p, \non\\
 & &\sigma_C^{\rm np}=\dfrac{2\sigma_0}{x_0}
        \left[\left(1-\dfrac{4}{x_0}-\dfrac{8}{x_0^2}\right)\ln(x_0+1)
        +\dfrac{1}{2}+\dfrac{8}{x_0}-\dfrac{1}{2(x_0+1)^2}\right]\ \non \\
 & &\sigma_p=\dfrac{2\sigma_0}{x_0}\left[\left(1+\dfrac{2}{x_0}\right)
     \ln(x_0+1)-\dfrac{5}{2}+\dfrac{1}{x_0+1}-\dfrac{1}{2(x_0+1)^2}\right].
\label{eq3}
\eeq

It is possible to tailor the shape of the energy distribution
by selecting a convenient polarization for the colliding electron and
laser beams: {\em flat} distributed scattered photons when laser and electrons 
have {\em like-handed} polarization, and quite {\em monochromatic} scattered 
photons (peaked at $y_{\rm max}$) by colliding {\em opposite-handed} electrons 
and photons \cite{NIM219}. 
The monochromaticity of the spectrum improves by increasing the energy
of the initial beams, but the annihilation of a laser photon with a high energy
backscattered photon into an $e^+e^-$ pair will occur above the threshold for 
this reaction, $W^2_{\gamma\gamma'}=(\omega_{\rm max}+\omega_0)^2-(
\omega_{\rm max}-\omega_0)^2>4m^2_e$. Therefore, one generally demands $x_0 
\lsim 2(1+\sqrt{2})\approx 4.83$ to prevent this unwanted process. 

\subsection{Two-photon processes}

The high energy backscattered photons produced from one electron/positron beam 
of the linear collider can be brought into collision with similarly produced
photons from the other beam, resulting in $\gamma\gamma$
processes. 

Because of the small, but non-zero, photon scattering
angles, the luminosity distribution depends on the conversion
distance (distance from the conversion point, where the laser pulse
intersects the electron beam, to the interaction point)
and on the size and shape the electron beam would have in the
absence of the laser. The effect of a non-zero conversion distance $b$
enters through a geometrical factor $\rho=b\theta_0/a$,
where $a$ is the radius of the electron beam, assumed round, at the interaction 
point.\footnote{Elliptical beams have been considered in \cite{ellip}.}
Typically, $b$ is of the order of some cm and $a$ is of the order of 100~nm. 
If $\rho\gg 1$, only the photons with higher
energy can meet at the i.p. and the spectrum becomes narrower. On
the contrary, when $\rho\ll 1$ all kind of photons can collide and yield
a broader luminosity spectrum. 

Other effects on the luminosity distribution will be neglected such 
as multiple scattering of electrons in the laser, deflection by an external
magnetic field (proposed to remove spent electrons from the interaction
region \cite{NIM294}), synchrotron radiation between the conversion and the 
interaction point, etc. \cite{NIM335}.

The cross section of the process $\gamma\gamma\to X$
with polarized photons can be written in the the Stokes-parameter and
in the photon-helicity bases as
\beq	
\dsh_{\gaga}=\sum_{i,j=0}^3\x_i\tx_j\dsh_{ij}=\sum_{a,b,c,d=\pm}
\r_{ac}\tilde{\r}_{bd}M_{ab}M^*_{cd}\dg,
\label{eq1}
\eeq
where $\x_i$, $\tx_j$ are the Stokes parameters of the first and
second photon,\footnote{
Tilded parameters refer to the second photon.} respectively; $\x_0=\tx_0=1$;  
$\r_{\pm\pm}=\frac{1}{2}(1\pm\x_2)$, $\r_{+-}=\r^*_{-+}=\frac{1}{2}
(-\x_3+i\x_1)$, and the same for $\tilde{\r}$,
are the photon polarization density matrices; $M_{ab}$ are the invariant
scattering amplitudes with photon helicities $a,b=\pm 1$; and
$\dg$ is the corresponding element of phase space divided by the
incoming flux.

Comparing both sides of Eq.~(\ref{eq1}), one finds sixteen independent
real functions $\dd\hat\sigma_{ij}$:
\beq
\frac{1}{4}\dg(|\mpp|^2+|\mmm|^2+|\mpm|^2+|\mmp|^2)&=&\dsh_{00}\equiv\dsh\non\\
\frac{1}{4}\dg(|\mpp|^2+|\mmm|^2-|\mpm|^2-|\mmp|^2)&=&\dsh_{22}\equiv
		\frac{1}{2}(\dsh_0-\dsh_2)\equiv\dd\tau^a\non\\
\frac{1}{4}\dg(|\mpp|^2-|\mmm|^2+|\mpm|^2-|\mmp|^2)&=&\dsh_{20}\non\\
\frac{1}{4}\dg(|\mpp|^2-|\mmm|^2-|\mpm|^2+|\mmp|^2)&=&\dsh_{02}\non\\
\dg\re(\mpp\mmm^*)&=&\dsh_{33}-\dsh_{11}\equiv\dsh_{||}-\dsh_\bot
\equiv\dd\tau\non\\
\dg\im(\mpp\mmm^*)&=&(\dsh_{13}+\dsh_{31})\non\\
\dg\mpm\mmp^*&=&(\dsh_{33}+\dsh_{11})+i(\dsh_{13}-\dsh_{31})\non\\
-\dg\mpp\mpm^*&=&(\dsh_{03}+\dsh_{23})+i(\dsh_{01}+\dsh_{21})\non\\
-\dg\mmm\mmp^*&=&(\dsh_{03}-\dsh_{23})-i(\dsh_{01}-\dsh_{21})\non\\
-\dg\mpp\mmp^*&=&(\dsh_{30}+\dsh_{32})+i(\dsh_{10}+\dsh_{12})\non\\
-\dg\mmm\mpm^*&=&(\dsh_{30}-\dsh_{32})-i(\dsh_{10}-\dsh_{12}) ,
\label{cs16}
\eeq
where $\dsh=\frac{1}{2}(\dsh_0+\dsh_2)=\frac{1}{2}(\dsh_{||}+\dsh_\bot)$ is the 
cross section for unpolarized photons, $\dsh_0\ (\dsh_2)$ are the cross sections 
for photons whose total helicity is 0 ($\pm 2$) and $\dsh_{||}\ (\dsh_\bot)$  
for photons with parallel (orthogonal) linear polarizations.

Actually $M_{ab}M^*_{cd}$ is a shorthand notation for $M_{ab;r_i}M^*_{cd;r_i}$ 
where $r_i$ labels the spin configuration of the particles in the final state 
$X$. In case that {\em Parity is conserved} in the process 
$\gaga\to X$,\footnote{
Parity is not conserved in Higgs-boson production and this simplification
will not be applied in that case. We will employ the simplified expressions 
only for the quark-pair production, which is a pure QED process at tree level.
}
\beq
M_{ab;r_i}M^*_{cd;r_i}=M_{-a-b;-r_i}M^*_{-c-d;-r_i}
\label{eq10}
\eeq
{\em and} the final state polarizations are not analyzed, i.e. we perform a
{\em sum over the final spins}, one has
\beq
|\mpp|^2&=&|\mmm|^2 \non\\
|\mpm|^2&=&|\mmp|^2 \non\\
\mpp\mmm^*&=&\mmm\mpp^* \non\\
\mpm\mmp^*&=&\mmp\mpm^* \non\\
\mpp\mpm^*&=&\mmm\mmp^* \non\\
\mmm\mpm^*&=&\mpp\mmp^* .
\eeq
We are then left with eight independent functions: $\dsh$, $\dd\hat\tau^a
\equiv\frac{1}{2}(\dsh_0-\dsh_2)$, $\dd\hat\tau\equiv\dsh_{||}-\dsh_\bot$ and 
five more that drop
if {\em azimuthal emission angles are integrated} \cite{NIM219}. Under all 
these assumptions, Eq.~(\ref{eq1}) is simplified to
\beq
\dsh_{\gaga}=\dsh+\x_2\tx_2\dd\hat\tau^a+
\frac{1}{2}(\x_3\tx_3-\x_1\tx_1)\dd\hat\tau .
\label{tote}
\eeq

Taking the colliding Compton-backscattered photons as the
initial state for the process $\gaga\to X$, the event rate can be written as
\beq
\dd N\equiv\dd L\ \langle\dsh_{\gaga}\rangle =
\dd L \sum_{i,j=0}^3\langle\x_i\tx_j
\rangle\dsh_{ij},
\label{avg}
\eeq
where $\dd L$ is the differential $\gaga$ {\em luminosity}  and 
$\langle\x_i\tx_j\rangle$ is the average of the product of the Stokes 
parameters along the interaction region. Only the diagonal products of 
Stokes parameters are relevant as may be seen from (\ref{tote}).

\section{The $\gaga$ luminosity}

Let $\pca$ ($\pta$) be the mean circular (linear) laser polarization 
and $\mbox{\boldmath{$\zeta_{||}$}}=2{\ple}$ ($\pte$) the mean longitudinal 
(transverse) polarization of the electron/positron beam. 
Of course, ${\pca}^2+{\pta}^2\le 1$, $(2{\ple})^2+{\pte}^2\le 1$, and the same 
for the polarizations of the second electron and laser beams.
In the Gaussian beam approximation \cite{NIM219}, the $\gaga$ luminosity reads
\beq
\dd L=L_{\rm eff}f_C(y)f_C(\tilde{y})I_0(v)\exp\left\{-\frac{\rho^2}{2}
\left(\frac{y_m}{y}+\frac{y_m}{\tilde{y}}-2\right)\right\}M\delta(y\tilde{y}-z^2)
\dd y\dd\tilde{y} ,
\eeq
where $z^2\equiv W^2/s_{e^+e^-}$, $W^2$ is the two-photon invariant mass,
$v\equiv\rho^2\sqrt{\left({y_{\rm max}}/{y}-1\right)
\left({y_{\rm max}}/{\tilde{y}}-1\right)}$ and $M\equiv 1+A_1
\langle\cos \Psi\rangle+A_2\langle\cos 2\Psi\rangle$, with $A_n(y,\tilde y)$
being functions proportional to ${\pta\tpta}$ \cite{NIM219}. The averages of 
azimuthal angles
are given by $\langle\cos n\Psi\rangle=I_n(v)/I_0(v)$ where $I_n$ are the 
modified Bessel functions of $n-$th order. 
For zero conversion distance $I_0(0)=1$ and $I_{n>0}(0)=0$. 
The total effective luminosity is $L_{\rm eff}=k^2 L_{e^+e^-}$ where $k^2$ is a
conversion coefficient.\footnote{Alhough $L_{\rm eff}<L_{e^+e^-}$,
the $\gaga$ luminosity can be larger than the $e^+e^-$ luminosity because of
the absence of beam-beam effects that allows a larger $L_{e^+e^-}$ 
\cite{NIM294}. On the other hand, for $\rho\ge0$, $\int \dd L\le L_{\rm eff}$ 
which produces an additional reduction in the total luminosity available.} 

In the Gaussian beam approximation the average of the Stokes-parameters
$\langle\x_i\tx_j\rangle$ can be obtained analytically as well \cite{NIM219}.
We employ here an average at a fixed invariant mass,
\beq
\langle\x_i\tx_j\rangle_z \equiv \dis\int_y\ \dd L\ \langle\x_i\tx_j\rangle 
\left/
	\dis\int_y\dd L
\right. ,
\eeq
which is of interest since the $\gaga\to X$ cross section does not depend 
on the rapidity of the two-photon system (equivalently the energy fraction $y$ 
of one of the laser photons).

To illustrate some polarization effects that will be exploited later,
we consider below two different polarizations for the lasers, assuming arbitrary 
longitudinal polarizations $\ple$, $\tple$ for the electron/positron beams.

\subsection{Circularly polarized lasers}

The relevant averages of Stokes parameters yield
\beq
\langle\x_2\tx_2\rangle&=&\langle\x_2\rangle\langle\tx_2\rangle=\x_2\tx_2 ,
\non\\
\langle\x_3\tx_3\rangle&=&-\langle\x_1\tx_1\rangle\equiv\Lambda.
\label{cir}
\eeq
$\Lambda$ vanishes for $\rho=0$ and has very small values otherwise, in 
particular when $y\sim\tilde{y}\sim y_{\rm max}$. 

\begin{figure}
\begin{center}
\begin{tabular}{c}
\epsfig{file=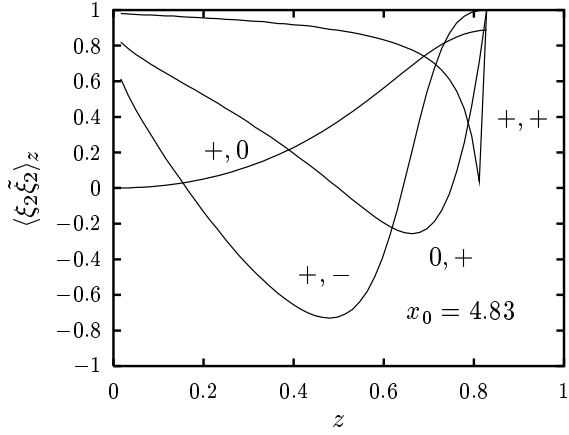,angle=0,width=0.45\linewidth}
\end{tabular}
\caption{
The average $\langle\x_2\tx_2\rangle_z$ for $x_0=4.83$ and several values
of the electron longitudinal and laser circular polarizations $({\ple},{\pca})$,
taking the same configuration for both arms of the collider: ${\ple}={\tple}$ 
and ${\pca}={\tpca}$. For the case ${\ple}=-{\tple}$ and ${\pca}=-{\tpca}$ all 
the curves change sign.
\label{stokescir}
}
\end{center}
\end{figure}

The average $\langle\x_2\tx_2\rangle_z$ (Fig.~\ref{stokescir}) is, on the 
contrary, independent of conversion distance effects for circularly polarized
photons. 
The event rate (\ref{avg}) can be very conveniently rewritten as 
\beq
\dd N&=&\dd L^{J_z=0}\ \dsh_0+ \dd L^{J_z=\pm2}\ \dsh_2 , 
\non\\
\dd L^{J_z=0}&\equiv&\frac{1}{2}\dd L\ (1+\langle\x_2\tx_2\rangle), 
\ \ 
\dd L^{J_z=\pm2}\equiv\frac{1}{2}\dd L\ (1-\langle\x_2\tx_2\rangle), 
\label{lumcir}
\eeq
where a contribution $\dd L\ \langle\x_3\tx_3-\x_1\tx_1\rangle$ has been 
safely neglected. 

\subsection{Linearly polarized lasers}

Neglecting $\rho\ne0$ effects, fully accounted for in our numerical calculations, 
one can write 
\beq
\langle\x_2\tx_2\rangle&\simeq&\langle\x_2\rangle\langle\tx_2\rangle \ =
 4{\ple}{\tple}\ {\cir}\tilde{\cir}
\non\\
\langle\x_3\tx_3-\x_1\tx_1\rangle&\simeq&\langle\x_3\rangle\langle\tx_3\rangle-
\langle\x_1\rangle\langle\tx_1\rangle=
{\pta}{\tpta}\ {\ell}\tilde{\ell}\cos2(\Delta\gamma) .
\label{lin}
\eeq
$\Delta\gamma$ is the angle between the planes of maximal linear 
polarization of both lasers.

\begin{figure}
\begin{center}
\begin{tabular}{c}
\epsfig{file=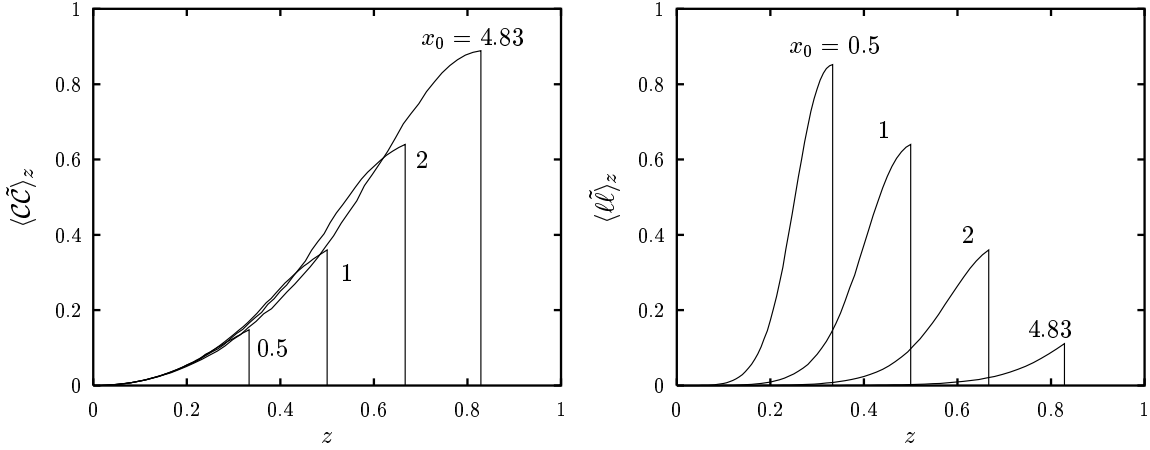,angle=0,width=0.9\linewidth}
\end{tabular}
\caption{
The correlations $\langle{\cir}\tilde{\cir}\rangle_z$ and 
$\langle{\ell}\tilde{\ell}\rangle_z$, relevant for linearly polarized laser
beams (\ref{lin}), for different values of $x_0$.
\label{stokes}
}
\end{center}
\end{figure}

Both the induced circular ${\cir}$ and linear ${\ell}$ polarizations
of the backscattered photons are growing functions with $y$ (Fig.~\ref{stokes}).
But, while the circular polarization is larger for high values
of $x_0$, the longitudinal one is enhanced for low values of $x_0$. 
For this configuration it results more convenient to cast (\ref{avg}) into
\beq
\dd N&=&\dd L^{||}\ \dsh_{||}+ \dd L^{\perp}\ \dsh_\perp + \dd L^C \ \dd\hat\tau^a, 
\non\\
\dd L^{||}&\equiv&\frac{1}{2}\dd L\ 
(1+\langle\x_3\tx_3-\x_1\tx_1\rangle), 
\ 
\dd L^{\perp}\equiv\frac{1}{2}\dd L\ 
(1-\langle\x_3\tx_3-\x_1\tx_1\rangle), \
\dd L^C\equiv\dd L\ \langle\x_2\tx_2\rangle.
\label{lumlin}
\eeq
The contribution $\dd L^C$ is roughly 
proportional to the product of the longitudinal polarizations of 
electron/positron beams (\ref{lin}) and plays an important role for
background suppression: since the $q\bar q$ background far from threshold has 
mostly $J_z=\pm2$, the term $\dd\tau^a$ is negative and the contribution 
proportional to $\dd L^C$ further suppresses the background by choosing 
like-handed polarized linac beams (${\ple}{\tple}>0$).

\subsection{Luminosity spectrum}
 
\begin{figure}
\begin{center}
\begin{tabular}{c}
\epsfig{file=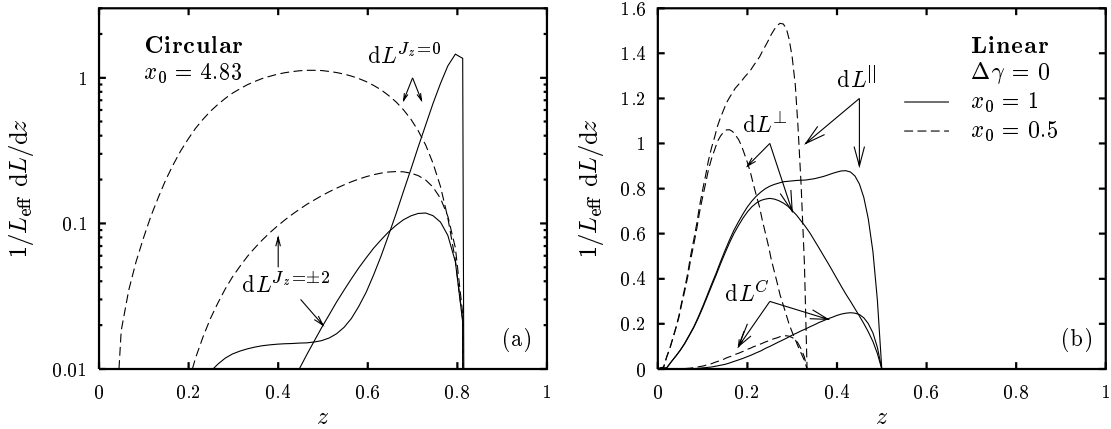,angle=0,width=0.9\linewidth}
\end{tabular}
\caption{
Normalized luminosities for circularly polarized lasers (a); 
and linearly polarized lasers with $\Delta\gamma=0$ (b). The lasers are assumed
to be completely polarized and the electrons 85\% longitudinally polarized. 
Identical configurations are taken for both arms of the collider. 
In (a) $x_0=4.83$ and solid (dashed) lines are for opposite-handed 
(like-handed) photons and electrons with $\rho=3.0$ ($\rho=0.6$). 
In (b) $\rho=0.6$, ${\ple}={\tple}=0.85/2$ and $x_0=1$ ($x_0=0.5$) for solid 
(dashed) lines.
\label{lumis}
}
\end{center}
\end{figure}

The resulting luminosity spectrum for the various components and different
polarizations is displayed in Fig.~\ref{lumis}. 
We assume realistic degrees of longitudinal
electron polarization (85\%) and laser polarizations (both circular and 
linear close to 100\% seem feasible \cite{kotser}). 

For circularly polarized lasers \cite{borden} (Fig.~\ref{lumis}a), a {\em broad}
spectrum (dashed lines) can be achieved by employing electrons and 
laser photons with like-handed helicities (see (+,+) in Fig.~\ref{stokescir}) 
and a small value of $\rho=0.6$
to allow low energetic backscattered photons in the interaction region.
On the contrary, a {\em sharp} spectrum (solid lines) peaking in the vicinity 
of $z=y_{\rm max}$
can be obtained using opposite-handed electrons and laser photons (see (+,$-$) 
in Fig.~\ref{stokescir}) in a more restrictive interaction
region $\rho=3.0$. The $J_z=0$ events (containing maybe the Higgs 
signal) are efficiently enhanced and the $J_z=\pm 2$ components (including the 
non-Higgs background) are suppressed by tuning lasers and linac 
beams so that $x_0=4.83$ (maximal profit of the collider energy). The peak takes 
place at $z_{\rm opt}\approx 0.8$. One can obtain Higgs bosons with 
$M_\varphi\lsim400$ (800) GeV at a linac energy $\sqrt{s_{e^+e^-}}=500$~GeV 
(1 TeV).

For linearly polarized lasers, a small $\rho=0.6$ has
been chosen. The case of parallel laser polarizations ($\Delta\gamma=0$) is 
shown in Fig.~\ref{lumis}b.
For $\Delta\gamma=\pi/2$ the roles of $\dd L^{||}$ and $\dd L^{\perp}$ get 
exchanged.
There is a compromise between getting a good separation of the $||$ and $\perp$ 
components (small $x_0$ is preferred) and producing a heavy enough Higgs boson, 
since the energy available is proportional to $x_0/(x_0+1)$.
The best solution consists of choosing the minimum possible value of $x_0$
and tuning the linac and laser energies to 
\beq
\sqrt{s_{e^+e^-}}&=&\frac{M_\varphi}{z_{\rm opt}} 
\label{tunedlinac}
\\
\omega_0	 &=&\frac{m^2_e}{2M_\varphi}x_0z_{\rm opt} \simeq
130\ x_0z_{\rm opt}\ \frac{\rm eV}{M_\varphi/{\rm GeV}}
\label{tunedlaser}
\eeq
so that the Higgs boson sits on an optimal 
\beq
z_{\rm opt}=\alpha(x_0)\frac{x_0}{x_0+1}
\eeq 
with $\alpha(x_0)\simeq 0.9$ for $x_0\lsim 1$ \cite{Gunion}.  
Such tunable linac and laser energies, $\sqrt{s_{e^+e^-}}\lsim 1$~TeV and
$\omega_0\sim 0.1\div1$~eV, are foreseen \cite{tesla}.
In this operation mode, assuming a maximal linac energy $\sqrt{s_{e^+e^-}}=
500$~GeV (1 TeV), one can produce Higgs bosons with masses up to
$M_\varphi=150$ (300) GeV for $x_0=0.5$, and 
$M_\varphi=225$ (450) GeV for $x_0=1.0$. Of course larger values of $x_0$
could be tuned to produce heavier Higgs bosons but then the distinction of 
their CP-parity becomes more difficult.

\section{Higgs-boson phenomenology at the PLC}

\subsection{Resonant production of Higgs bosons}

The production of a spin-0 CP-eigenstate $\varphi$ by the fusion of two photons,
$\gaga\to\varphi$, proceeds via $F_{\mu\nu}F^{\mu\nu}\varphi$ for a scalar 
field, and $F_{\mu\nu}\tilde{F}^{\mu\nu}\varphi$ for a pseudoscalar. The 
corresponding Feynman rules for such couplings are proportional to $\epsilon
\cdot\tilde\epsilon$ and $(\epsilon\times\tilde\epsilon)_z$, respectively, for 
two back-to-back photons moving along the $z$-axis.
The amplitude for a general spin-0 state coupled to two photons can be
written as
\beq
M_{\lambda\tilde{\lambda}}=(\epsilon\cdot\tilde\epsilon)\ {\cal E}
            		  +(\epsilon\times\tilde\epsilon)_z\ {\cal O},
\label{cp2}
\eeq
where ${\cal E}$ (${\cal O}$) are the CP-even(odd) contributions to the
amplitude. CP is violated if {\em both} terms are not vanishing. In any case the 
coupling for opposite helicity photons vanishes ($\mpm=\mmp=0$). There are 
four independent functions describing the process, out of the sixteen in 
(\ref{cs16}), 
\beq
\dsh_{00}+\dsh_{22}&=&\frac{1}{2}\dg(|\mpp|^2+|\mmm|^2)= 
|{\cal E}|^2+|{\cal O}|^2\non\\
\dsh_{20}+\dsh_{02}&=&\frac{1}{2}\dg(|\mpp|^2-|\mmm|^2)= 
-2\im({\cal E}{\cal O}^*)\non\\
\dsh_{31}+\dsh_{13}&=&\dg\im(\mpp\mmm^*)= -2\re({\cal E}{\cal O}^*) \non\\
\dsh_{33}-\dsh_{11}&=&\dg\re(\mpp\mmm^*)= |{\cal E}|^2-|{\cal O}|^2.
\eeq
Defining the asymmetries \cite{Grzad}:
\beq
{\cal A}_1&\equiv&\frac{|\mpp|^2-|\mmm|^2}{|\mpp|^2+|\mmm|^2}
=-\frac{2\im({\cal E}{\cal O}^*)}{|{\cal E}|^2+|{\cal O}|^2} \non\\
{\cal A}_2&\equiv&\frac{2\im(\mpp\mmm^*)}{|\mpp|^2+|\mmm|^2}
=-\frac{2\re({\cal E}{\cal O}^*)}{|{\cal E}|^2+|{\cal O}|^2} \non\\
{\cal A}_3&\equiv&\frac{2\re(\mpp\mmm^*)}{|\mpp|^2+|\mmm|^2}
=\frac{|{\cal E}|^2-|{\cal O}|^2}{|{\cal E}|^2+|{\cal O}|^2},
\label{asymmetries}
\eeq
the event rate (\ref{avg}) is
\beq
\dd N&=& \dd L^{S=0}\ \dsh ,\non\\
\dd L^{S=0}&\equiv&\dd L \ [\ 1+\langle\x_2\tx_2\rangle+\langle\x_2+\tx_2\rangle
\ {\cal A}_1
 +\langle\x_3\tx_1+\x_1\tx_3\rangle\ {\cal A}_2
 +\langle\x_3\tx_3-\x_1\tx_1\rangle\ {\cal A}_3\ ] ,
\label{eq38}
\eeq 
with the unpolarized $\gaga\to\varphi$ cross section being $\dsh=\frac{1}{4}\dg
(|\mpp|^2+|\mmm|^2)$. Recalling the remarks above (\ref{tote}), expression 
(\ref{eq38}) shows that one has to analyze the spins and/or the azimuthal 
emission angles of the final state to probe CP violation through the 
asymmetries ${\cal A}_1$ and ${\cal A}_2$.

For $\varphi$ being a CP-eigenstate with $\eta^\varphi_{\rm CP}=\pm 1$ for $\varphi=H,A$
(scalar or pseudoscalar, respectively), one has ${\cal A}_1={\cal A}_2=0$, 
${\cal A}_3=\eta^\varphi_{\rm CP}$ and (\ref{eq38}) simplifies to
\beq
\dd L^{S=0}=\dd L^{0^{\eta^\varphi_{\rm CP}}}\equiv\dd L\ [\ 1+\langle\x_2\tx_2\rangle
+\eta^\varphi_{\rm CP}\langle\x_3\tx_3-\x_1\tx_1\rangle\ ].
\label{eq39}
\eeq
In fact, if CP is conserved, $\mpp=\eta^\varphi_{\rm CP}\mmm$ and,
since $\dg\re(\mpp\mmm^*)=\dsh_{||}-\dsh_\bot=\eta^\varphi_{\rm CP}\cdot
(\dsh_{||}+\dsh_\bot)$, 
only photons with parallel (orthogonal) linear polarizations couple to scalars 
(pseudoscalars), in agreement with (\ref{cp2}). We concentrate on this case
(CP conservation) in our analysis. It corresponds to the MSSM with real 
parameters.\footnote{
Actually, some of the parameters of the MSSM can be complex. In this case, 
physical phases may be introduced that produce CP-violating effects, through
the one-loop coupling of the Higgs field to two photons. 
A detailed study of the photon collider capabilities to probe CP violation in 
the MSSM Higgs sector of the MSSM has been recently presented in \cite{choi}.}
Notice that {\em only if the lasers are linearly polarized} it is then 
possible to distinguish the CP nature of the Higgs field, since
the average $\langle\x_3\tx_3-\x_1\tx_1\rangle$ is negligible for circularly
polarized lasers (\ref{cir}). 

\subsection{The $H$ and $A$ decays}

The Higgs spectrum of the MSSM is determined at tree level by two independent 
parameters, the pseudoscalar mass $M_A$ and the ratio of the two vacuum 
expectation values $\tan\beta$.
For the evaluation of the Higgs boson decays we have employed an
adapted version of the program {\tt HDECAY} \cite{hdecay}. We have
explored two scenarios, low and high $\tan\beta$ ($\tan\beta=2,50$,
respectively) and taken a reference set of MSSM input parameters
for which the decays of the neutral Higgs bosons into supersymmetric
particles are kinematically suppressed or forbidden in the studied
range of Higgs boson masses: $\mu=-250$~GeV, $M_2=m_{\tilde Q}=1$~TeV
and both no-mixing and maximal mixing in the squark sector, $A_t=0$ 
and $A_t=\sqrt{6} m_{\tilde Q}$, respectively. We present
here only the case of no-mixing. The squark mixing does not play a relevant 
role for the heavy Higgs sector. GUT gaugino-mass constraints 
and universal supersymmetric soft-breaking terms are assumed.
The main decay channels of $H$ and $A$ are the following \cite{decays}.
In the low $\tan\beta$ region, one must distinguish two Higgs mass
ranges: below the top-pair threshold, $H$ decays most often to light 
Higgs-pairs $hh$ (or to $WW^*$ for $M_H\lsim 200$~GeV if there is maximal 
mixing) and $A$ goes to $b\bar b$ ($M_A\lsim 200$~GeV) and $hZ$ 
(otherwise); whereas $H$ and $A$ decay fully to $t\bar t$ for 
$M_\varphi\gsim 2m_t$. If $\tan\beta$ is large, both $H$ and $A$ 
decay almost fully into $b\bar b$ pairs.

The background to these Higgs signals comes mainly from the continuum 
(non-Higgs) production of the same final states. The processes $\gaga\to hh$ 
\cite{hh} and $\gaga\to Zh$ are considered as negligible backgrounds 
\cite{Gunion}.

\subsection{Effective cross sections}

Since the process under study proceeds through a  narrow resonance,
one must take into account the detector accuracy when 
comparing signal (resonant production) and background (continuum production).  
An appropriate way to obtain {\em effective} signal and background consists of
integrating a window of two-photon invariant masses $(M_\varphi-\Delta,
M_\varphi+\Delta)$ around the Higgs mass assuming the 
reconstruction resolution is a Gaussian \cite{borden}. That is, making
use of (\ref{avg}),  
\beq
N_{\rm eff}(M_\varphi)
&\equiv&L_{\rm eff}\dd\sigma_{\rm eff}\equiv
\int^{M_\varphi+\Delta}_{M_\varphi-\Delta}
\dd N_{\rm eff}(W),\non\\
\dd N_{\rm eff}(W)
&\equiv&\int^{y_{\rm max}\sqrt{s_{e^+e^-}}}_{M_X}
\ \dfrac{\dd W'}{\sqrt{2\pi}\delta}
\exp\left\{-\dfrac{(W'-W)^2}{2\delta^2}\right\}
\dfrac{\dd L}{\dd W'}\ \langle\dd\hat\sigma_{\gamma\gamma}\rangle.
\label{eff} 
\eeq
The signal cross section $\gamma\gamma\to \varphi\to X$ for 
polarized photons, ${\hat\sigma}^{\rm signal}_{\gaga}$, is {\em isotropic} and 
can be written as
\beq
{\hat\sigma}^{\rm signal}_{\gaga}(W)=8\pi\dfrac{\Gamma(\varphi\to\gamma\gamma)
\Gamma(\varphi\to X)}
{(W^2-M^2_\varphi)^2+\Gamma^2_\varphi M^2_\varphi}(1+\lambda\tilde\lambda).
\label{sg}
\eeq
For the regions where the width is much smaller than the detector resolution
the following relation can be used\footnote{
The Breit-Wigner distribution 
$\dfrac{1}{\pi}\dfrac{\Gamma M}{(W^2-M^2)+\Gamma^2M^2}\simeq
\delta(W^2-M^2)=\dfrac{1}{2M}\delta(W-M)$ for $\Gamma\ll M$.}
\beq
N^{\rm signal}_{\rm eff} (M_\varphi) &\simeq& R(\Delta/\delta) \
\left.\dfrac{\dd L^{S=0}}{\dd W}\right|_{W=M_\varphi} \times
4\pi^2
\dfrac{\Gamma(\varphi\to\gamma\gamma){\rm BR}(\varphi\to X)}{M^2_\varphi},
\label{signal}
\eeq
where the Gaussian error function describing the fraction of signal events 
contained in the bin $M_\varphi\pm\Delta$ is $R(\Delta/\delta)=0.9545$ 
for $\Delta=2\delta$ (standard deviations)
and $\dd L^{S=0}$ is the $\gaga$ luminosity for the production of a spin-0
state (\ref{eq38}). Assuming a resolution of $\delta=2.5$ GeV, it is safe
to use approximation (\ref{signal}) only for an intermediate Higgs boson mass 
(standard or supersymmetric) but it overestimates the production rates for 
heavier Higgs bosons. 
The MSSM Higgs widths can be $\Gamma_\varphi\approx 5$ GeV for 
$M_\varphi\lsim 500$ GeV (and much larger for the SM Higgs boson) though 
$\Gamma_\varphi\lsim 0.1$ GeV for masses below the top-pair threshold 
\cite{decays}.

The effective background $\gaga\to X$ for an invariant mass $W=M_\varphi$ is 
approximated by
\beq
N^{\rm bckg}_{\rm eff} (W) &\simeq& 2 \Delta\
\dfrac{\dd L}{\dd W}\ \langle\dsh^{\rm bckg}_{\gamma\gamma}(W)\rangle ,
\label{bkg}
\eeq
assuming the distribution of invariant masses weighted with the luminosity 
spectrum is smooth enough. The main source of continuum background to the
Higgs boson production is the quark-pair production.
The total Born cross sections $\gaga\to q\bar q$ ($\hat\sigma_0$, 
$\hat\sigma_2$ and $\hat\tau\equiv\hat\sigma_{||}-\hat\sigma_\perp$) for
a polar cut in the two-photon center of mass system $|\cos\vartheta^*|<c$ are
\beq
\hat\sigma_0(W)&=&\dfrac{12\pi\alpha^2 Q^4_q}{W^2}(1-\beta^4)
\left\{\dfrac{1}{2}\ln\dfrac{1+c\beta}{1-c\beta}+\dfrac{c\beta}{1-(c\beta)^2}
\right\}
\label{sigma0}
\\
\hat\sigma_2(W)&=&\dfrac{12\pi\alpha^2 Q^4_q}{W^2}
\left\{\dfrac{5-\beta^4}{2}\ln\dfrac{1+c\beta}{1-c\beta}
-c\beta\left[2+\dfrac{(1-\beta^2)(3-\beta^2)}{1-(c\beta)^2}
\right]\right\}
\label{sigma2}
\\
\hat\tau(W)=\hat\sigma_{||}-\hat\sigma_{\perp}
&=&\dfrac{12\pi\alpha^2 Q^4_q}{W^2}(1-\beta^2)^2
\left\{\dfrac{1}{2}\ln\dfrac{1+c\beta}{1-c\beta}+\dfrac{c\beta}{1-(c\beta)^2}
\right\}< 0 \,
\label{par-perbckg}
\eeq
with $\beta=\sqrt{1-4m^2_q/W^2}$ and $Q_q$ the electric charge of the 
quark $q$.
We take the usual $c=0.7$ that helps
to eliminate a great deal of background events in comparison to only a 30\% 
reduction of signal events. 
In practice this means a cut on the difference of rapidities of both jets/tops 
given by $|\Delta\eta|\le|\ln[(1+\beta c)/(1-\beta c)]|$.

The background cross sections receive important QCD radiative corrections 
\cite{qcd}, particularly large for the $J_z=0$ component, and also significant 
electroweak corrections \cite{ew}. The QCD corrections to the
interference term (\ref{par-perbckg}) are not yet available \cite{avto}. 
In this preliminary analysis we ignore them all: the statistical significances
of the Higgs signals should not change dramatically and the CP asymetries
will not suffer a large distortion due to the background as long as 
$\hat\sigma_{||}$ and $\hat\sigma_{\perp}$ are corrected by similar amounts.
However, all relevant QCD and electroweak corrections to the Higgs production 
and decay are incorporated in {\tt HDECAY} \cite{hdecay,decays}.

There is of course no interference between the $H$ and $A$ amplitudes, owing to
their opposite CP-parity, even if their masses are very close, as it is usually 
the case, since we consider {\em unpolarized} final states.
We neglect the interference between the resonant production and the 
continuum background, although it may be relevant because of the not
so narrow width of the Higgs bosons, particularly for very high 
masses.\footnote{
See Ref.~\cite{asakawa} for a very recent analysis of the interferences
between $H$ and $A$ with small mass gap and between them and the continuum
for polarized top-pairs in the final state. There it is proposed an
alternative method to measure the CP-parity of the Higgs bosons based on
these interferences. It appears not so efficient as the one analyzed here.}

\begin{figure}
\begin{center}
\begin{tabular}{c}
\epsfig{file=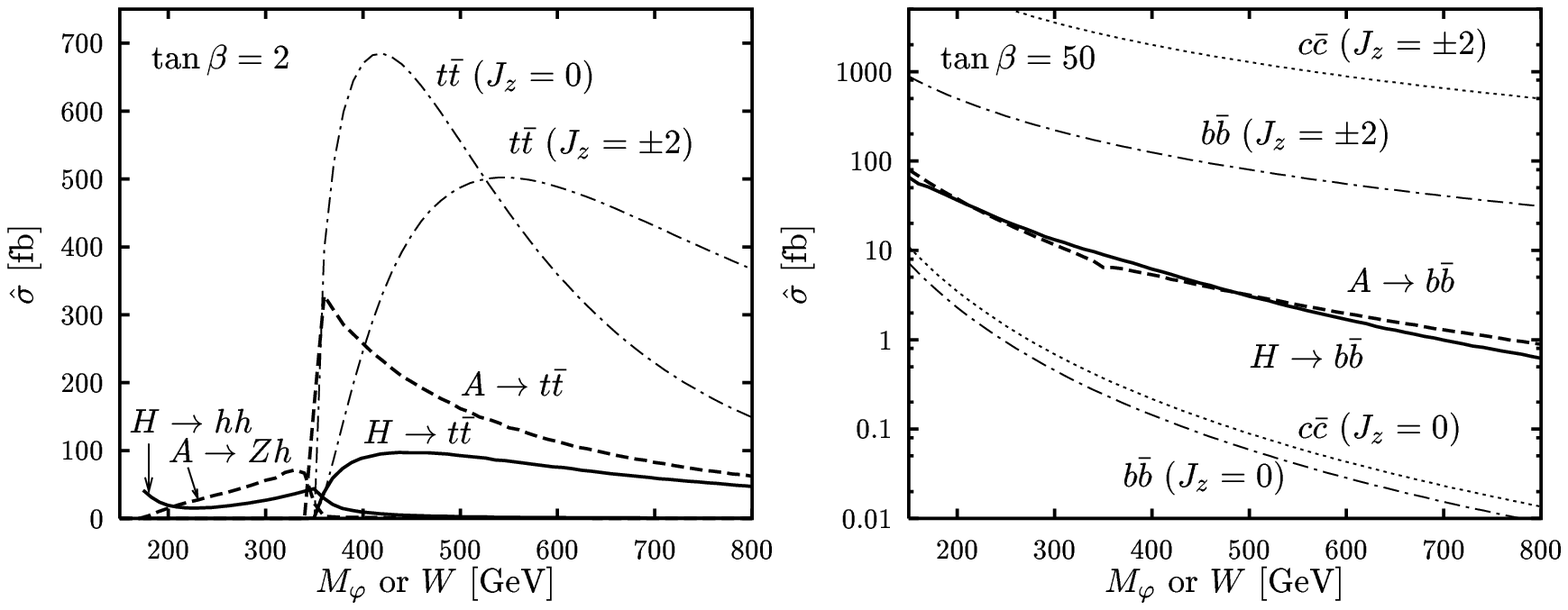,angle=0,width=0.9\linewidth}\\
\epsfig{file=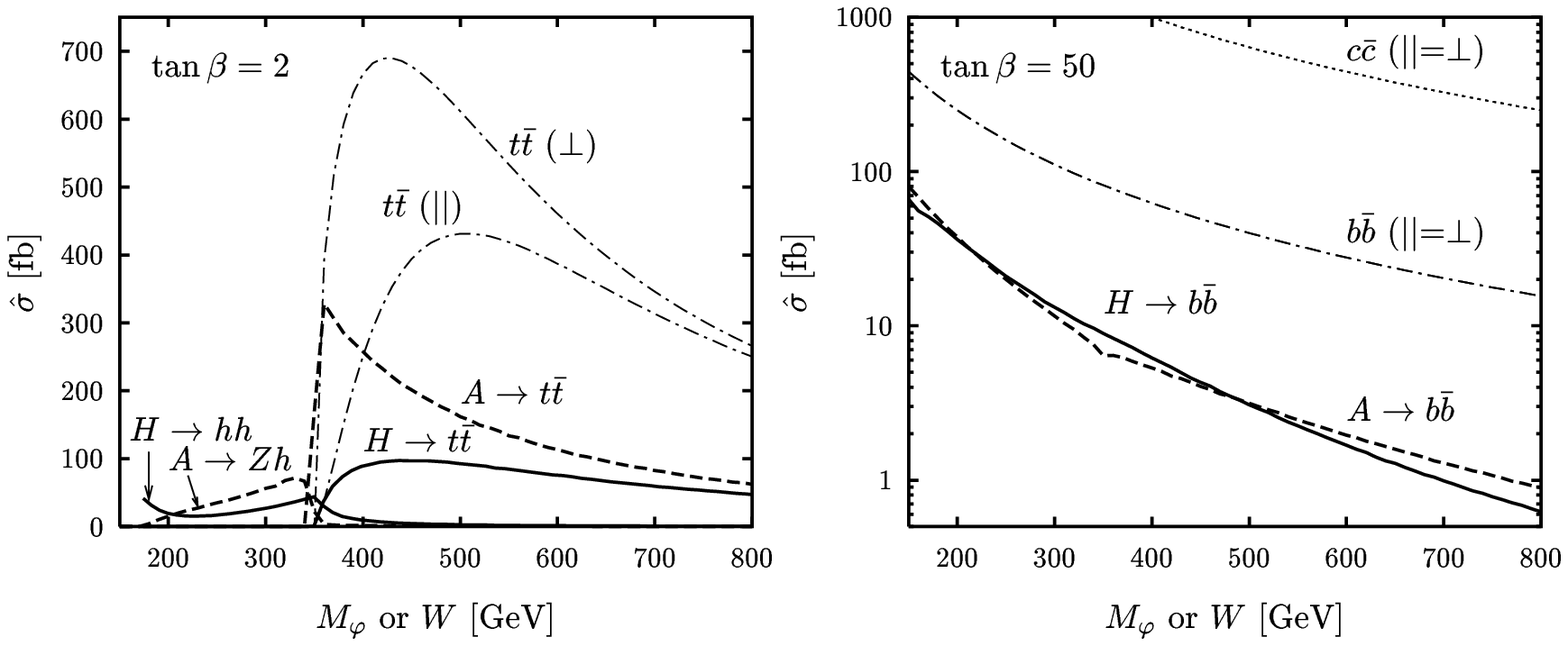,angle=0,width=0.9\linewidth}
\end{tabular}
\caption{
The Born cross sections ($|\cos\vartheta^*|<0.7$) for relevant signals and 
backgrounds,
before folding with the $\gaga$ luminosity, for two scenarios of $\tan\beta$.
They are split into initial states of $J_z=0$ and $J_z=\pm2$ (above), and
parallel ($||$) and perpendicular ($\perp$) photon linear polarizations (below). 
They are normalized as explained in the text.
\label{xsect}
}
\end{center}
\end{figure}

In Fig.~\ref{xsect} we show the cross sections (before 
folding with the $\gaga$ luminosity) for $\gaga\to X$ and $\gaga\to H/A\to X$. 
The resonant ones are normalized assuming the events collected in an interval 
of invariant masses around $W=M_\varphi$ with a width $2\Delta=10$ GeV.
The $q\bar q$ events in the continuum and far from threshold ($b\bar b$ and
$c\bar c$ pairs) are mainly in the $J_z=\pm 2$ state. They 
are produced by photons with perpendicular and parallel polarizations at
the same rate.
Near threshold ($t\bar t$ pairs) the $J_z=0$ component dominates and the 
quark-pairs couple mostly to perpendicularly polarized photons. 

\section{The case of circularly polarized lasers}

Circularly polarized lasers can produce the largest luminosities 
(Fig.~\ref{lumis}a) for the Higgs signal (\ref{eq39}) and, at the same time, 
the $J_z=0$ component of the continuum background (\ref{lumcir}). The latter
is suppressed far from the $q\bar q$ threshold (Fig.~\ref{xsect}).
The broad spectrum in Fig.~\ref{lumis}a (dashed lines)
for a fixed collider energy is clearly optimal for discovery; the sharp one 
(solid lines), for a collider energy tuned at the peak, is the best to reduce 
the large $J_z=\pm2$ component of the background. The same is 
true for an intermediate Higgs mass \cite{borden}. 
The determination of the CP-parity of the Higgs field is not possible using 
circularly polarized lasers, as already mentioned.

\begin{figure}
\begin{center}
\begin{tabular}{c}
\epsfig{file=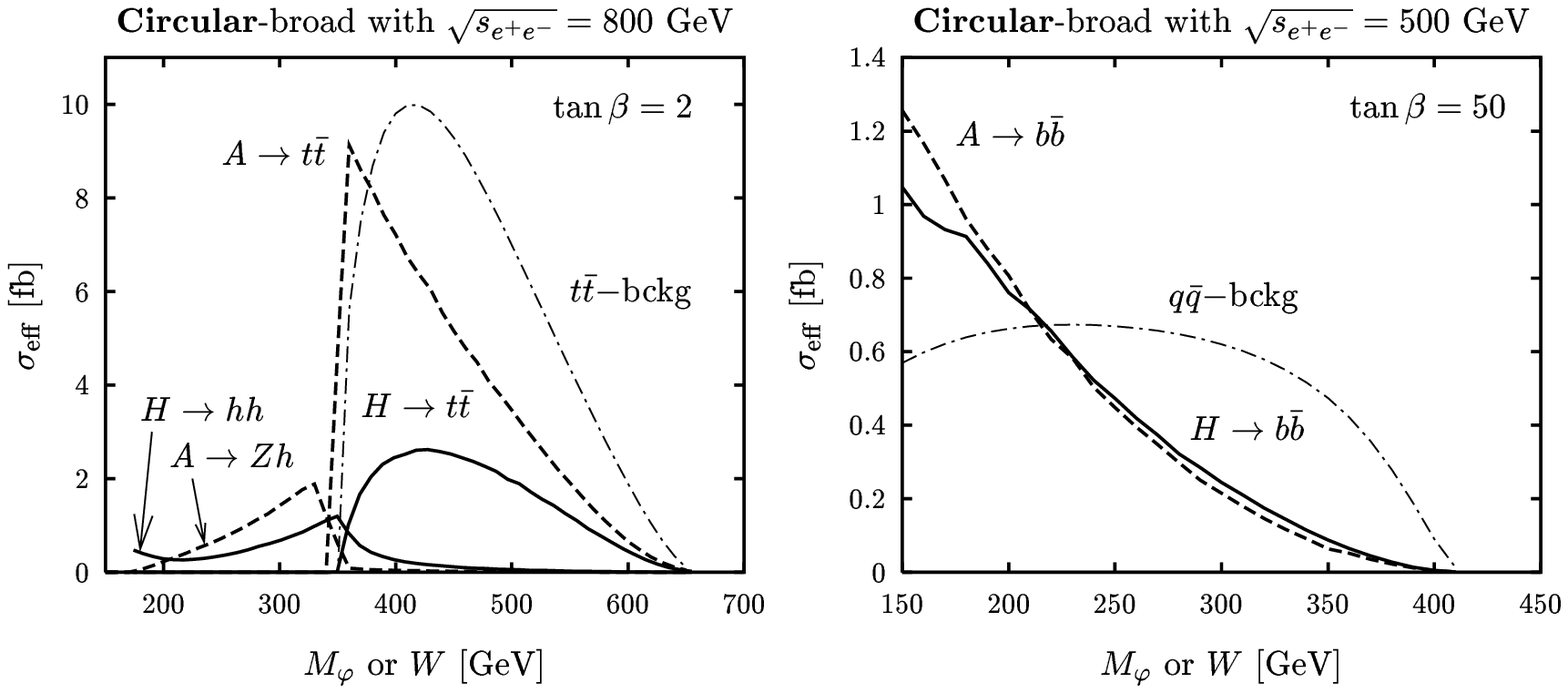,angle=0,width=0.9\linewidth}\\  
\epsfig{file=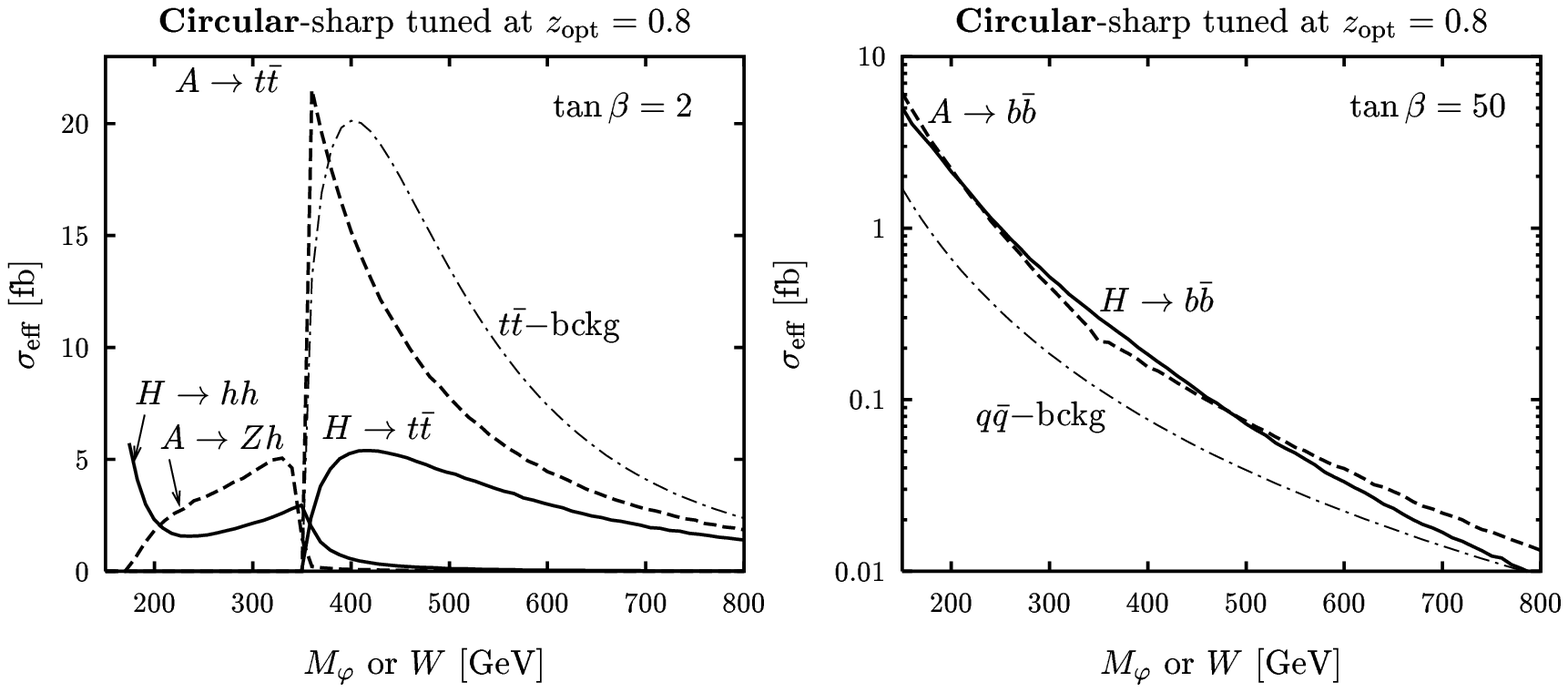,angle=0,width=0.9\linewidth}
\end{tabular}
\caption{The effective cross sections for the $H$ and $A$ signals and their
backgrounds in two $\tan\beta$ scenarios and for two modes of operation
of the PLC, both with circular laser polarizations. 50\% $b\bar b$-tagging
efficiency and 5\% $c\bar c/b \bar b$ contamination have been assumed.
\label{fig:effsigmacir}
}
\end{center}
\end{figure}

These expectations are illustrated in Fig.~\ref{fig:effsigmacir}, where
the $\gaga$ luminosities of Fig.~\ref{lumis}a have been assumed.
For the $b\bar b$ channel we have considered a 50\% $b\bar b$-tagging 
efficiency and the possibility of a  5\% $c\bar c/b \bar b$ contamination.
The $t\bar t$ pairs are supposed to be fully identified. The $hh$ and $Zh$ 
channels are taken free of background.

For low $\tan\beta$ the $A$ signal dominates over $H$, but in the 
large $\tan\beta$ region both $H$ and $A$ are produced with similar rates.
This is because the production rates are roughly proportional to the two-photon 
decay width of the corresponding Higgs boson (\ref{sg},\ref{signal}), 
which is much larger for the pseudoscalar if $\tan\beta$ is small 
\cite{gagawidth}.
On the other hand the production rates are smaller for high $\tan\beta$,
in particular for large Higgs masses, due to the smaller two-photon widths 
when $\tan\beta$ is large and to the smearing of the resosonance
(the total Higgs-boson widths grow with $\tan\beta$).

For the $t\bar t$ channel the background is large because near threshold
the pairs produced in the continuum have a large $J_z=0$ component. Conversely,
the background suppression is very effective in the $b\bar b$ channel, 
especially for the sharp spectrum running in the tuned-energy mode.

In the fixed-energy mode, we have taken $\sqrt{s_{e^+e^-}}=800$ GeV in order
to better 
explore the low $\tan\beta$ region, dominated by the $t\bar t$ channel above 
threshold, and $\sqrt{s_{e^+e^-}}=500$ GeV for the high $\tan\beta$ scenario 
where the Higgs bosons  decay into $b\bar b$ pairs.
Larger statistics are obtained for the sharp spectrum in the tuned-energy mode
($\sqrt{s}=M_\varphi/z_{\rm opt}$).  

\begin{figure}
\begin{center}
\begin{tabular}{c}
\epsfig{file=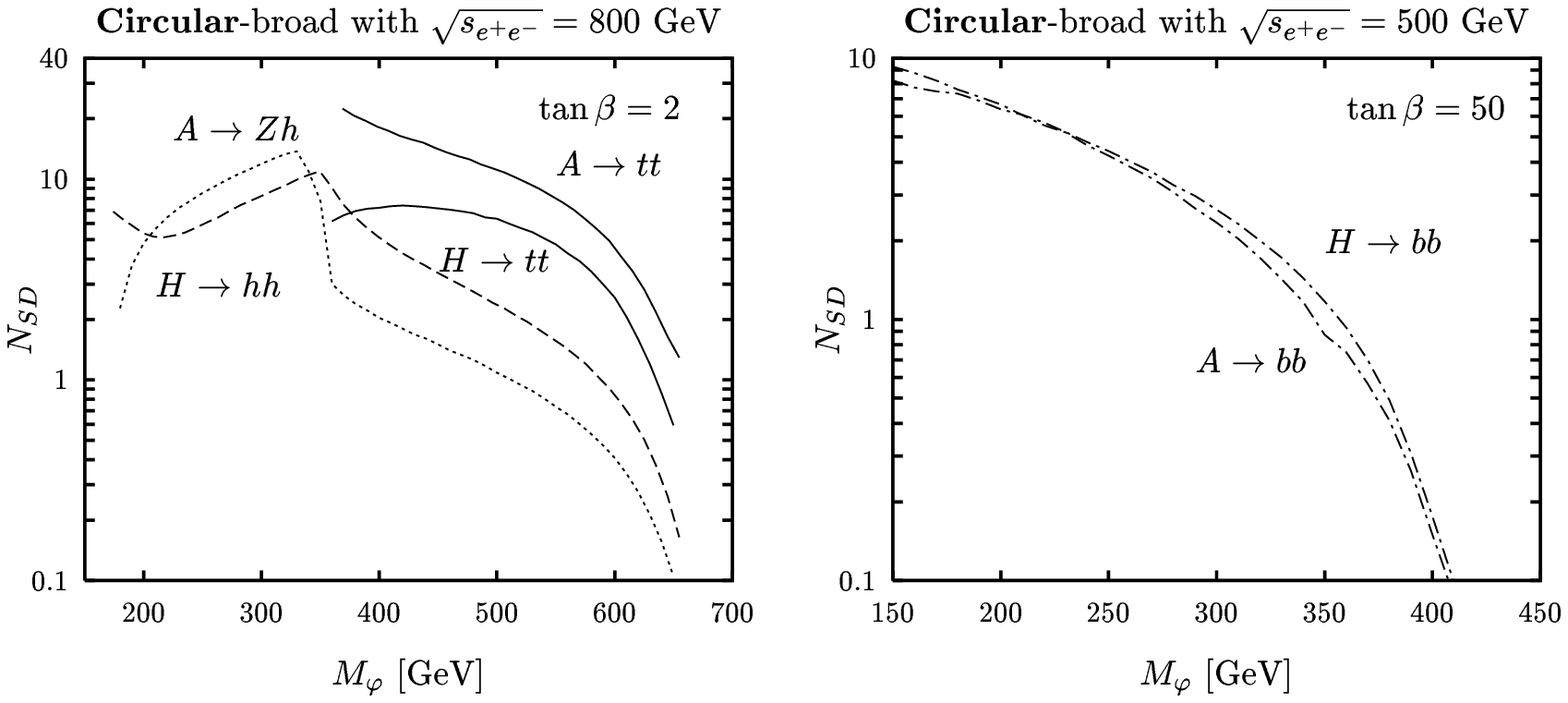,angle=0,width=0.9\linewidth}\\
\epsfig{file=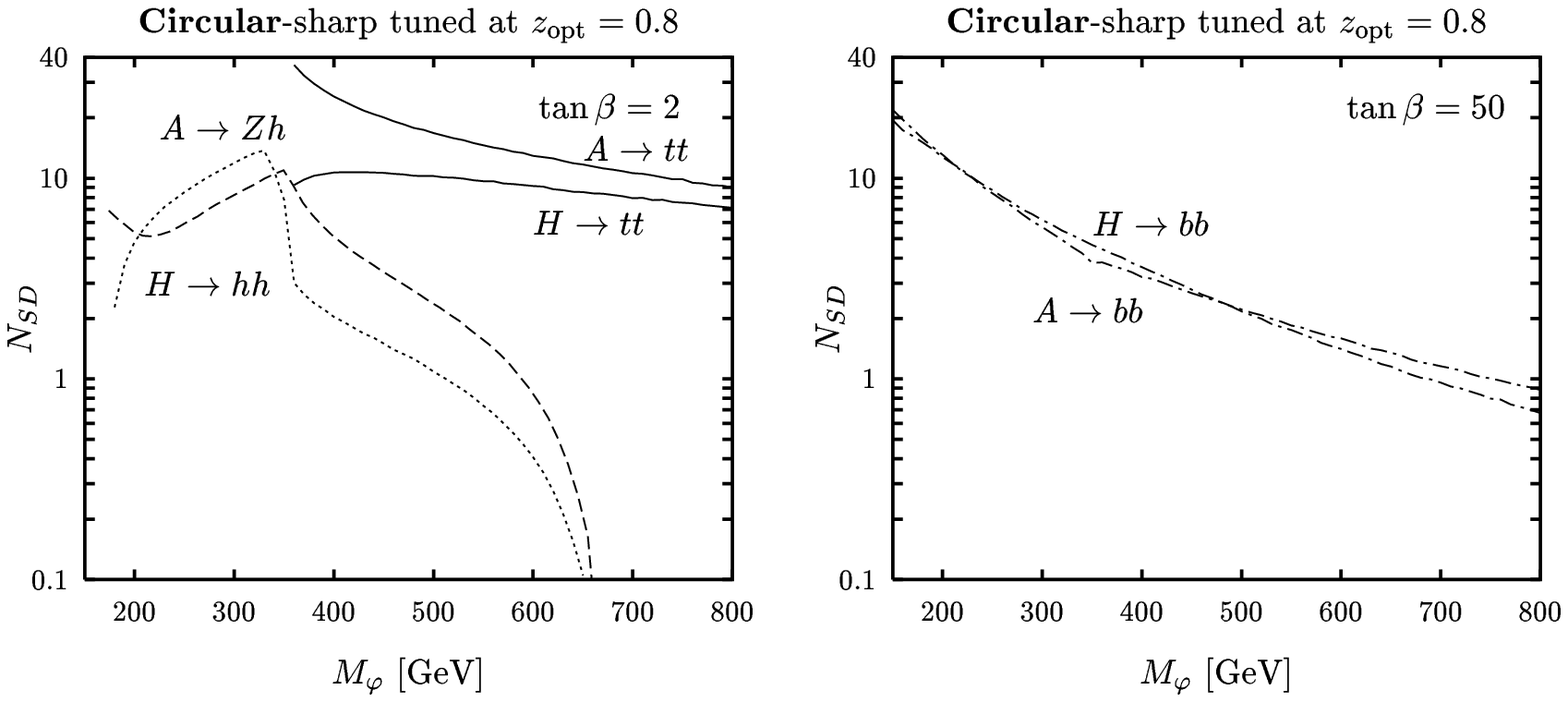,angle=0,width=0.9\linewidth}
\end{tabular}
\caption{Statistical significance, $N_{SD}$, of the $H$ and $A$ signals 
for $L_{\rm eff}=100\mbox{ fb}^{-1}$ in the two modes of operation of the PLC 
with circular laser polarizations, for $t\bar t$ (solid), $b\bar b$
(dot-dashed), $hh$ (dashed) and $Zh$ (dotted) final states.
\label{fig:statsigncir}
}
\end{center}
\end{figure}

The statistical significances for the Higgs signals, defined by
\beq
N_{SD}(\varphi)=\frac{\sigma^\varphi_{\rm eff}}{\sqrt{\sigma^\varphi_{\rm eff}+
\sigma^{\rm bckg}_{\rm eff}}}\ \sqrt{L_{\rm eff}},
\eeq
are shown in Fig.~\ref{fig:statsigncir}, assuming $L_{\rm eff}=100\mbox{
fb}^{-1}$. 
A range of both $H$ and $A$ Higgs masses up to the kinematical limit
$M_\varphi\approx 0.8\sqrt{s_{e^+e^-}}$ can be covered with statistical 
significances of several standard deviations, particularly for
not very large masses, in both $\tan\beta$ scenarios, even in the fixed-energy 
mode.

\section{The case of linearly polarized lasers}

Linearly polarized lasers (Fig.~\ref{lumis}b) are necessary to distinguish 
the CP-parity of the Higgs bosons. The asymmetry ${\cal A}_3$ in 
(\ref{asymmetries})
is probed by taking the difference of the event rates with $\Delta\gamma=0$ 
(lasers with parallel polarizations) and with $\Delta\gamma=\pi/2$ 
(lasers with perpendicular polarizations) \cite{Gunion,Grzad,Khuen},
\beq
\langle{\cal A}^\varphi_3\rangle=\frac{\sigma_{\rm eff}(\Delta\gamma=0)-
\sigma_{\rm eff}(\Delta\gamma=\pi/2)}
{\sigma_{\rm eff}(\Delta\gamma=0)+\sigma_{\rm eff}(\Delta\gamma=\pi/2)},
\label{a3}
\eeq
where the contamination from a possible background is included in 
$\sigma_{\rm eff}\equiv\sigma^\varphi_{\rm eff}+\sigma^{\rm bckg}_{\rm eff}$.
In terms of the linac and laser polarizations this asymmetry reads
\beq
\langle{\cal A}^\varphi_3\rangle\simeq{\pta\tpta}\langle\ell\tilde\ell
\rangle_{z_{\rm opt}}\frac{\eta^\varphi_{\rm CP}\hat\sigma^\varphi
+\frac{1}{2}(\hat\sigma^{\rm bckg}_{||}-\hat\sigma^{\rm bckg}_{\perp})}
{\frac{1}{2}(1+4{\ple\tple}\langle{\cir}\tilde{\cir}\rangle_{z_{\rm opt}})
(2\hat\sigma^\varphi+\hat\sigma^{\rm bckg}_0)
+\frac{1}{2}(1-4{\ple\tple}\langle{\cir}\tilde{\cir}\rangle_{z_{\rm opt}})
\hat\sigma^{\rm bckg}_2}.
\eeq
The $\rho\ne0$ effects have been dropped in the previous equation but are
included in the numerical calculations. This expression explains
the importance of linear laser polarizations, ${\pta}$, ${\tpta}$, the role of
the linac longitudinal polarizations, ${\ple}$, ${\tple}$, and the effect
of the different components of the background. The induced polarizations of 
the Compton photons are shown in Fig.~\ref{stokes}. We take fully polarized
lasers, ${\ple}={\tple}=0.85/2$ and a realistic interaction region with 
$\rho=0.6$. 

\begin{figure}
\begin{center}
\begin{tabular}{c}
\epsfig{file=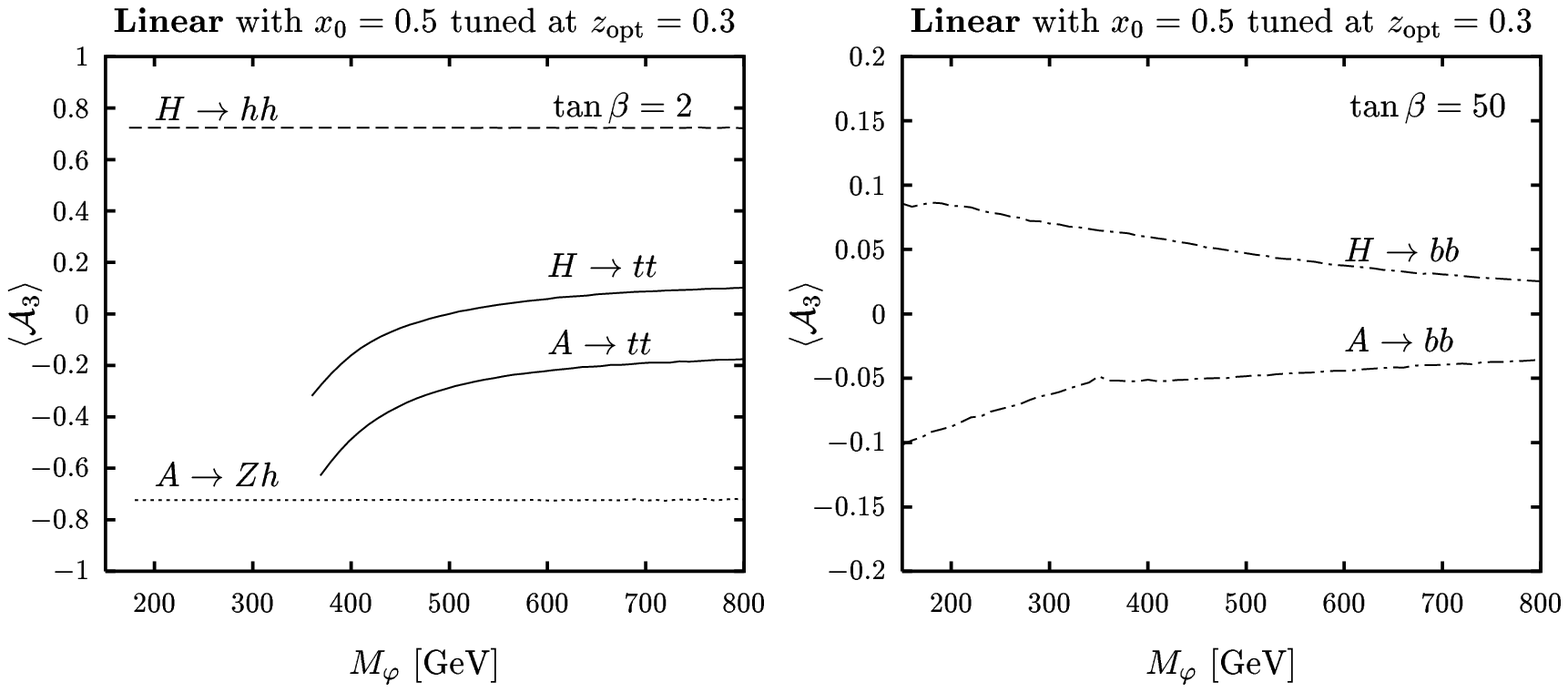,angle=0,width=0.9\linewidth}\\
\epsfig{file=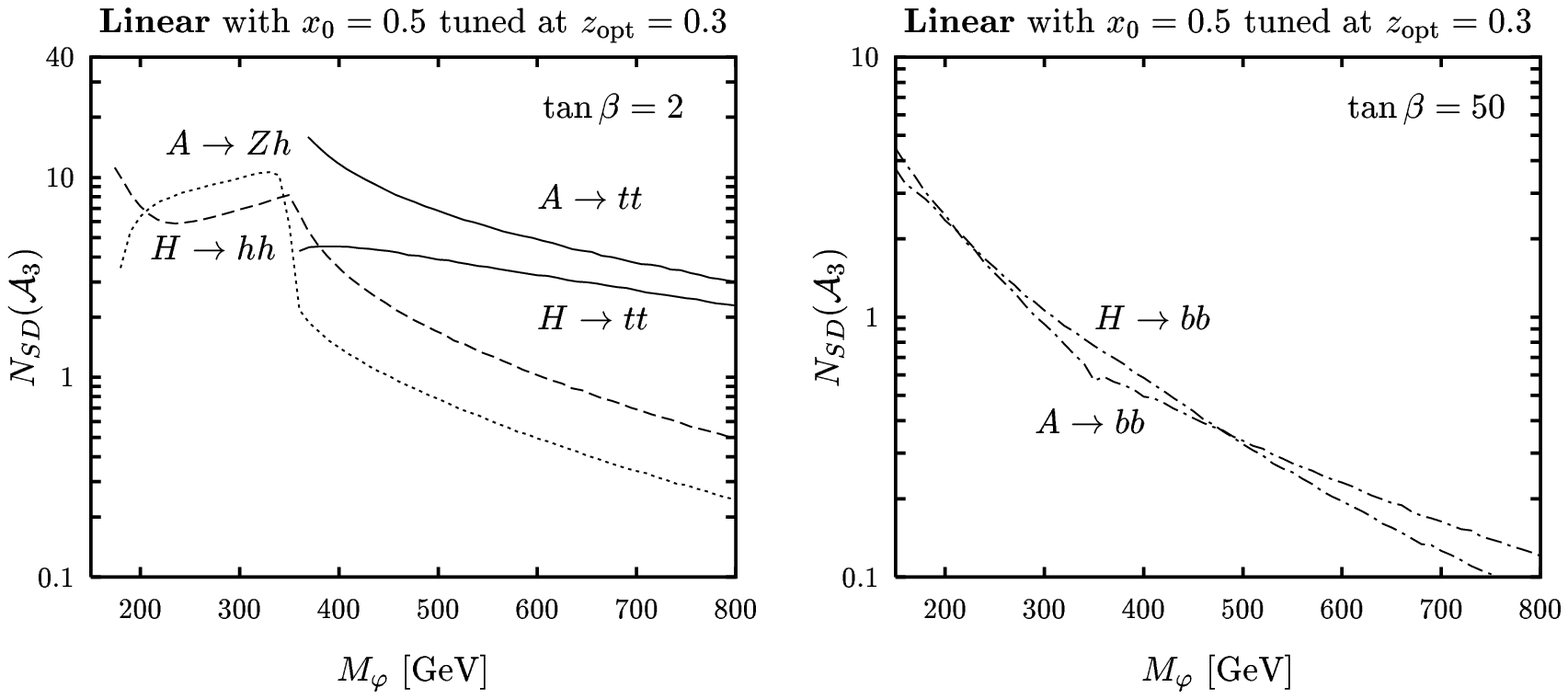,angle=0,width=0.9\linewidth}
\end{tabular}
\caption{The expectation value of the asymmetry ${\cal A}_3$ and its
statistical significance ($L_{\rm eff}=100\mbox{ fb}^{-1}$) for the PLC 
operating with linear polarizations at a tuned energy with $x_0=0.5$.
The final states are labelled as before: $t\bar t$ (solid), $b\bar b$
(dot-dashed), $hh$ (dashed) and $Zh$ (dotted).
\label{fig:a3x0=0.5}
}
\end{center}
\end{figure}

\begin{figure}
\begin{center}
\begin{tabular}{c}
\epsfig{file=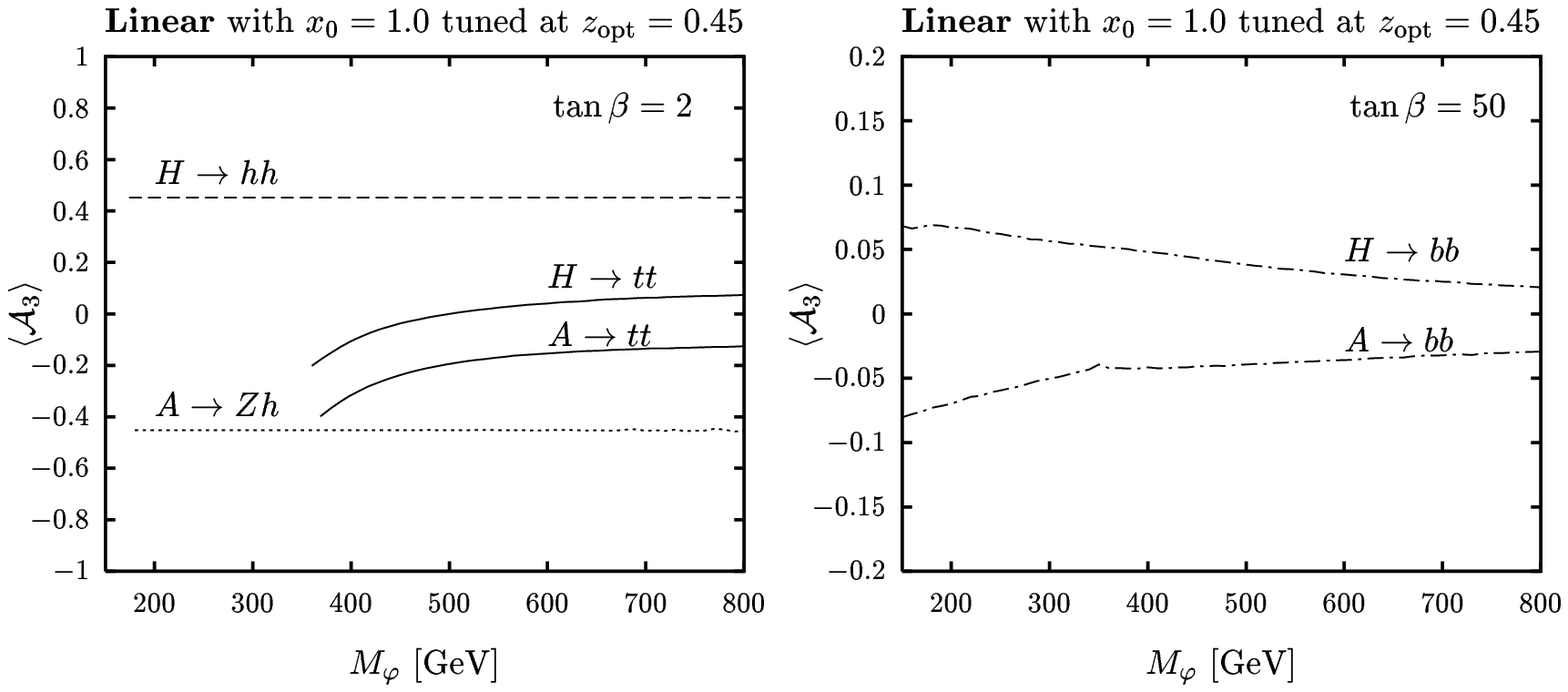,angle=0,width=0.9\linewidth}\\
\epsfig{file=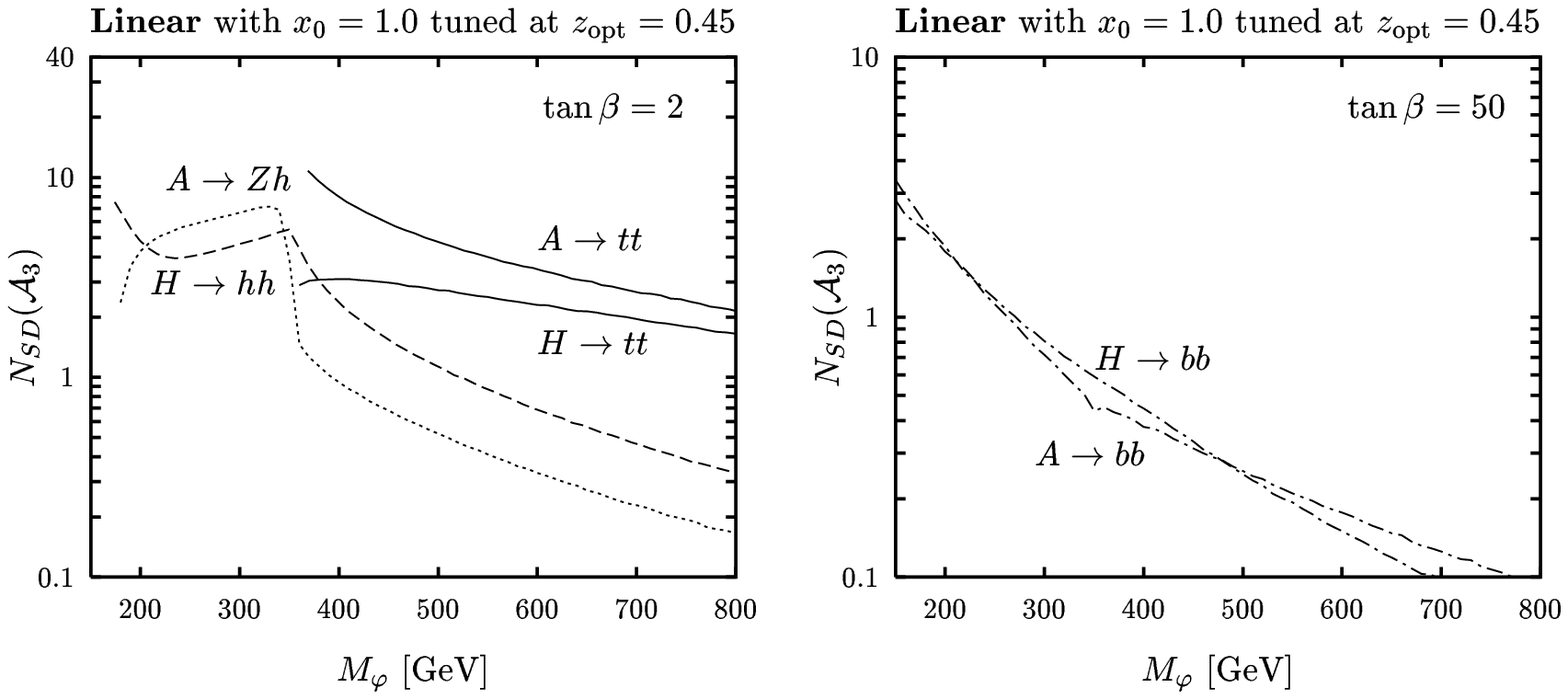,angle=0,width=0.9\linewidth}
\end{tabular}
\caption{The same as in Fig.~\ref{fig:a3x0=0.5} but with $x_0=1.0$
\label{fig:a3x0=1.0}
}
\end{center}
\end{figure}

For the channels $H\to hh$ and $A\to Zh$, considered to be free of background, 
the asymmetries are maximal: $\langle {\cal A}^\varphi_3 \rangle=
\eta^\varphi_{\rm CP}\cdot 0.72$ (0.45) for $x_0=0.5$ (1.0), respectively 
(Figs.~\ref{fig:a3x0=0.5}, \ref{fig:a3x0=1.0}).

Notice that $\langle{\cal A}_3\rangle$ does not vanish in absence of a signal. 
Furthermore,
the background contribution to the asymmetry from $q\bar q$ final states is 
always {\em negative} (\ref{par-perbckg}), since $\hat\tau=\hat\sigma_{||}-
\hat\sigma{_\perp}<0$, although it is negligible very far from threshold
($b\bar b$ and $c\bar c$ pairs). For the $t\bar t$ channel the situation is 
different (see Fig.\ref{xsect}) and ${\cal A}_3$ has even ``wrong" sign for
$M_H\lsim 500$ GeV. In any case, the presence of background
reduces the absolute value of the asymmetry due to the increase of the
denominator of (\ref{a3}). The use of like-handed linac beams helps
to somewhat control the typically large $J_z=\pm2$ component of the background,
but the necessary choice of a small $x_0$ makes it less efficient in 
comparison to the case of maximal $x_0=4.83$ and circularly polarized lasers, 
discussed in the previous Section.

The statistical significance of the CP-asymmetry is given by \cite{Gunion}
\beq
N_{SD}({\cal A}^\varphi_3)=\frac{|\sigma^\varphi_{\rm eff}(\Delta\gamma=0)-
\sigma^\varphi_{\rm eff}(\Delta\gamma=\pi/2)|}
{\sqrt{\sigma_{\rm eff}(\Delta\gamma=0)+\sigma_{\rm eff}(\Delta\gamma=\pi/2)}}
\ \sqrt{L_{\rm eff}}.
\eeq
Two runs with different laser polarizations are necessary. We take, 
for definiteness, a somewhat optimistic $L_{\rm eff}=100\mbox{ pb}^{-1}$
per run. 
The linac and laser energies are tuned for these analyses so that
the Higgs bosons sit at $z_{\rm opt}$ for a given value of $x_0$. We presume
that the Higgs mass(es) will be known by some other means, so that 
such tuning is possible, and that the laser and linac can reach the demanded 
energies available (see Eqs.~\ref{tunedlinac}, \ref{tunedlaser}). 
Taking $x_0=1.0$ one can get 50\% larger Higgs boson masses than with
$x_0=0.5$, but then smaller asymmetries are obtained. Nevertheless,
the statistical significances do not improve very dramatically for the
smaller $x_0$ (compare Fig.~\ref{fig:a3x0=0.5} with Fig.~\ref{fig:a3x0=1.0}). 

A clear distinction of the CP-parity of the heavy MSSM Higgs bosons is possible
for the low $\tan\beta$ scenario in the 
whole range of accessible masses of the photon collider, based on tunable
linac and lasers ($M_\varphi\le$ 450 GeV for $\sqrt{s_{e^+e^-}}\le$ 1 TeV).
Below the top-pair threshold, the CP-even (-odd) Higgs bosons decay into
light-Higgs pairs $hh$ (or $Zh$, respectively), and for heavier Higgs bosons,
into top pairs. The signal, particularly for the pseudoscalar, is very clear
in the low $\tan\beta$ scenario.
If $\tan\beta$ happens to be large, the $bb$ decay channel is favoured, and
the CP-asymmetry allows to distinguish the CP-parity only if the Higgs
bosons are lighter than $\sim300$ GeV, since the background-suppression power 
of this collider configuration is not so efficient as with circularly
polarized lasers. The signals are not so large as in the small $\tan\beta$ case
either.

Testing the CP-parity of $h$ is also feasible. A light Higgs boson decays
mostly into $b\bar b$ pairs. One can choose a small $x_0=0.5$ and tune
the collider at $z_{\rm opt}=0.45$ to cover the whole range of possible
masses [$M_h\lsim 100$ (130) GeV for low (high) $\tan\beta$],
for which a maximal $\sqrt{s_{e^+e^-}}\sim 250$ GeV is needed.
Statistical significances for the CP asymmetry of more than one standard 
deviation are obtainable \cite{Gunion}.

\section{Conclusions}

We have presented a self-contained introduction to the phenomenology
of a high energy photon collider, including a derivation of the $\gaga$ 
luminosity spectra with especial emphasis on the polarization effects and 
their influence on the resonant Higgs production and decay, as well as on
the most relevant backgrounds.

We have explored the photon-collider potential to observe and test
the CP-parity of the heavy neutral MSSM Higgs bosons.

With circularly polarized lasers one can maximally profit from the linear 
collider energy and, at the same time, very efficiently reduce the $J_z=\pm2$ 
component of the continuum background, by choosing an extreme value of
$x_0\equiv4\omega_0 E_b/m^2_e\approx2(1+\sqrt{2})$. For fixed linac and laser 
energies ($E_b$ and $\omega_0$, respectively) and like-handed lasers and
electrons or, even better, using opposite-handed electrons and lasers 
for both arms of the collider and tuning the linac energy to the luminosity
peak, heavy Higgs bosons with masses below 800 GeV can be observed
in a linear collider tunable up to $2E_b=1$ TeV. 
The tuned-energy option, but at more moderate energies, is also the best 
configuration to produce an intermediate-mass Higgs boson and measure its 
two-photon width with very good accuracy \cite{soldner}.

In the low $\tan\beta$ scenario, and below the top-pair threshold,
$H$ can be observed to decay into $hh$ and $A$ into $Zh$, but for heavier Higgs
bosons the $t\bar t$ final state is best, with a larger rate for the
pseudoscalar. In the high $\tan\beta$ scenario the $b\bar b$ channel
is common to both at a similar, but smaller rate than for low $\tan\beta$.

To determine the CP-parity of the Higgs bosons, laser linear polarizations
are absolutely necessary. The CP-even (CP-odd) Higgs bosons couple only
to two photons with parallel (orthogonal) polarizations. Samples rich
in those initial states can be prepared by choosing a smaller $x_0$ at
the price of a smaller Higgs-mass reach $M_\varphi\propto x_0/(x_0+1)$.
A longitudinal polarization of the linac beams helps to suppress the
background, although less efficiently than in the case of circular laser
polarizations. An asymmetry based on the comparison of production rates
for parallel and perpendicular laser-polarization planes allows a clear
distinction of $H$ and $A$. The CP-parity of Higgs bosons with masses
below $\sim$450 (300) GeV can be tested in the low (high) $\tan\beta$ region 
with a photon collider based on a linac tunable up to 1 TeV.

\noindent
{\bf Acknowledgements}

\noindent
I am very grateful to A. Djouadi for suggestions and comments, to A. Schiller 
and V. Serbo for a helpful communication and to T. Riemann and A. Tkabladze 
for discussions. 
I also thank A. Djouadi, M. Jack, T. Riemann and A. Tkabladze 
for carefully reading the manuscript.
This work has been partially supported by the Spanish CICYT and Junta de 
Andaluc{\'\i}a, under contracts AEN96-1672 and FQM101, respectively.


\end{document}